\documentclass[manuscript]{aastex}

\begin{document}

\title{Spectroscopy of V341 Arae: A Nearby Nova-like Variable inside a Bow-Shock
Nebula\altaffilmark{*}}

\author{
Howard E. Bond\altaffilmark{1,2,3}
and
Brent Miszalski\altaffilmark{4,5}
}

\altaffiltext{*}
{Based in part on observations with the 1.5-m telescope operated by the SMARTS
Consortium at Cerro Tololo Interamerican Observatory. Based in part on
observations collected at the European Organisation for Astronomical Research in
the Southern Hemisphere under ESO programme 073.D-0157(A).} 

\altaffiltext{1}
{Department of Astronomy \& Astrophysics, Pennsylvania State
University, University Park, PA 16802, USA; heb11@psu.edu}

\altaffiltext{2}
{Space Telescope Science Institute, 
3700 San Martin Dr.,
Baltimore, MD 21218, USA}

\altaffiltext{3}
{Visiting astronomer, Kitt Peak National Observatory, National Optical
Astronomy Observatory, which is operated by the Association of Universities for
Research in Astronomy (AURA) under a cooperative agreement with the National
Science Foundation.  
}

\altaffiltext{4} 
{South African Astronomical Observatory, PO Box 9, Observatory 7935, South
Africa}

\altaffiltext{5}
{Southern African Large Telescope Foundation, PO Box 9, Observatory 7935, South
Africa}

\begin{abstract}

V341~Arae is a 10th-magnitude variable star in the southern hemisphere,
discovered over a century ago by Henrietta Leavitt but relatively little studied
since then. Although historically considered to be a Cepheid, it is actually
blue and coincides with an X-ray source. The star lies near the edge of the
large, faint H$\alpha$ nebula Fr~2-11, discovered by D.~Frew, who showed that
V341~Ara is actually a cataclysmic variable (CV)\null. His deep imaging of the
nebula revealed a bow-shock morphology in the immediate vicinity of the star. We
have carried out spectroscopic monitoring of V341~Ara, and we confirm that it is
a nova-like CV, with an orbital period of 0.15216~days (3.652~hr). We show that
V341~Ara is remarkably similar to the previously known BZ~Cam, a nova-like CV
with a nearly identical orbital period, associated with the bow-shock nebula
EGB~4. Archival sky-survey photometry shows that V341~Ara normally varies
between $V\simeq10.5$ and 11, with a characteristic timescale ranging from about
10 to 16 days. V341~Ara lies well off-center within Fr~2-11. We speculate either
that the star is undergoing a chance high-speed encounter with a small
interstellar cloud, or that the nebula was ejected from the star itself in a
nova outburst in the fairly distant past. At a distance of only 156~pc, V341~Ara
is one of the nearest and brightest known nova-like variables, and we encourage
further studies.

\end{abstract}



\section{The Variable Star V341 Arae}

Variability of the 10th-magnitude star CPD\,$-63^\circ$4037, designated HV~2969
and later V341~Arae, was discovered at the Harvard College Observatory by
Henrietta Leavitt and announced by Pickering (1907). Over the ensuing century
the star attracted little attention. It was noted as being blue by Hoffmeister
(1956), who reported variability at a period of 11.95~days and considered
V341~Ara to be a Type~II Cepheid. The blue color was verified in a small number
of photometric observations during a study of Type~II Cepheids by Harris (1981).
He suggested it may instead be an eclipsing binary. 



Berdnikov \& Szabados (1998) obtained extensive photoelectric photometry and
also examined photometric data from the {\it Hipparcos\/} mission. They found a
period of 14.11~days, significantly longer than reported by Hoffmeister, and
with considerable photometric scatter; they also verified that the $V-I$ color
is relatively blue. Drilling \& Bergeron (1995), based on low-dispersion
objective-prism plate material, reported the star to have an OB spectral type,
which is inconsistent with the Cepheid classification. Further evidence that it
is not a Cepheid came from it being listed as an X-ray source, 1RXS
J165743.7$-$631237, in the {\it ROSAT\/} All-Sky Source Catalogue (Voges et al.\
1999). This led Kiraga (2012) to suggest that V341~Ara is most likely a
cataclysmic variable (CV)\null. Based on photometric monitoring by the All-Sky
Automatic Survey (ASAS; Pojma\'nski 2002), Kiraga found a modulation in its
brightness at a characteristic timescale  of about 10~days.

Our interest in V341~Ara arose from the discovery by David Frew that the star is
associated with a faint H$\alpha$ emission nebula, to which he gave the
designation Fr~2-11. He found the large ($8'\times6'$) nebula during a visual
inspection of images from the Southern H$\alpha$ Sky Survey Atlas (SHASSA;
Gaustad et al.\ 2001), aimed at discovering new planetary nebulae (PNe) at more
than $10^\circ$ from the Galactic plane. (Fr~2-11 lies at $b=-12\fdg5$.) The
H$\alpha$ nebula is actually also faintly visible in the red image of the
Digitized Sky Survey\footnote{Available from the Space Telescope Science
Institute at \url{\tt http://stdatu.stsci.edu/cgi-bin/dss\_form}}, as shown in
our Figure~1. However, it had not been recognized or cataloged prior to Frew's
discovery. Remarkably, V341~Ara is very far off-center in the faint nebula,
lying near its southwestern edge. 

Frew's discovery was communicated to us privately, and was presented at a
conference on PNe (Frew et al.\ 2006).  Further details were given in his PhD
dissertation (Frew 2008, hereafter F08), and Fr~2-11 was discussed briefly in
Frew et al.\ (2016). Of particular interest to us was F08's spectroscopic
confirmation that V341~Ara is indeed a nova-like CV, and that his follow-up
higher-resolution imaging had revealed a bow-shock morphology in the immediate
vicinity of the star, distinct from the large, low-surface-brightness H$\alpha$
nebula.

Emission nebulae associated with CVs that have not had a known nova outburst,
especially having a bow-shock structure, are extremely rare. The best-known
previous case is probably the parabolic-shaped nebula EGB~4, discovered by
Ellis, Grayson, \& Bond (1984, hereafter EGB)\null. EGB found the nebula to be
associated with a bright, previously unknown CV, subsequently designated BZ~Cam.
The EGB~4 nebula was discussed by Hollis et al.\ (1992), who argued that it
represents a case of a source with a strong wind passing supersonically through
a relatively stationary local interstellar medium (ISM)\null. BZ~Cam itself is a
nova-like variable (i.e., belonging to the ``UX~UMa'' subclass of CVs in which
there is a high rate of mass transfer from a main-sequence donor to its
white-dwarf companion, producing an optically thick accretion disk around the
white dwarf). This has led other authors (e.g., Griffith et al.\ 1995) to
suggest that the nebula was ejected from BZ~Cam itself and is being sculpted by
an interaction with the ISM\null. The orbital period of BZ~Cam is 0.1535~days
(3.68~hours; e.g., Patterson et al.\ 1996; Honeycutt et al.\ 2013). Deep optical
images of EGB~4 have been presented by Greiner et al.\ (2001). 

More recently the nebula IPHASX J210204.7+471015 has been shown to have
a bow-shock structure, which is associated with a previously unknown nova-like
CV (Guerrero et al.\ 2018). The central star has an orbital period of 4.26~hr.

Because of our interest in binary stars associated with PNe, we undertook
spectroscopic observations aimed at finding the orbital period and other
properties of V341~Ara, which we report in this paper.


\section{The Bow-Shock Nebula and {\em Gaia} Astrometry}


In order to illustrate the morphology of the bow shock associated with V341~Ara,
we retrieved a publicly available [\ion{O}{3}] image from the European Southern
Observatory (ESO) archive\footnote{The archive is located at \url{{\tt
http://archive.eso.org/eso/eso\_archive\_main.html}} and the data for Fr~2-11
were originally obtained as part of program 073.D-0157(A) (PI: Q.~A. Parker)}.
This observation was obtained with the Wide-Field Imager of the 2.2-m MPG/ESO
telescope on 2004 May~27. The exposure time was 900~s and employed a narrow-band
filter centered on the [\ion{O}{3}] 5007~\AA\ emission line. We cleaned the
image of cosmic rays using \textsc{lacosmic} (van~Dokkum 2001), and bad columns
were interpolated over. A false-color rendition of this image is shown in the
left panel of Figure~2. (F08 contains deeper images of the bow shock and
surrounding nebula, together with an extensive discussion of V341~Ara itself.)

For comparison, we show in the right-hand panel of Figure~2 a rendition of an
image of the EGB~4/BZ~Cam nebula, obtained by H.E.B. in 1996 with the Kitt Peak
National Observatory Mayall 4-m telescope. This false-color image combines
H$\alpha$ (red) and [\ion{O}{3}] 5007~\AA\ (green) exposures. 

Precise parallaxes and proper motions for both of these CVs are available from
the recent {\it Gaia\/} Data Release~2 (Gaia Collaboration et al.\ 2018). The
derived distances for V341~Ara and BZ~Cam, respectively, are $156.1\pm2.0$~pc
and $372.0\pm5.2$~pc. The respective proper motions are $97.7\,\rm mas\,yr^{-1}$
at position angle $200\fdg6$, and $28.8\,\rm mas\,yr^{-1}$ at PA $184\fdg6$.
These correspond to moderately large transverse velocities of 72 and $51\,\rm
km\,s^{-1}$, respectively. Both nebulae are remarkably similar in showing bright
rims of enhanced [\ion{O}{3}] emission that are almost exactly in the directions
of the stars' motions relative to the ISM\null. Based on average $V$ magnitudes
of 10.75 for V341~Ara (see below) and 12.8 for BZ~Cam (Honeycutt et al.\ 2013),
the stars' absolute magnitudes are also very similar, at $M_V=+4.8$ and +4.9,
respectively. These values are quite typical of nova-like CVs, and several
magnitudes brighter than the values for dwarf novae in quiescence (e.g., Ramsay
et al.\ 2017 and references therein).



%

\section{Light Curve Behavior}

For a star as bright as V341~Ara, there is now extensive photometric time-series
coverage available from all-sky monitoring surveys such as ASAS and the
follow-up All-Sky Automated Survey for Supernovae (ASAS-SN; Shappee et al.\ 2014
and Kochanek et al.\ 2017)\footnote{\url{\tt
http://www.astronomy.ohio-state.edu/asassn/index.shtml}}. Both the ASAS
data\footnote{\url{\tt http://www.astrouw.edu.pl/cgi-asas/asas\_variable\slash
165742-6312.6,asas3,0,0,500,0,0}} covering 2001 to 2009 and the ASAS-SN data
from 2016 to the present show that the star generally varied between
$V\simeq10.5$ and 11.0, with occasional brief excursions to somewhat fainter
levels. There have not been any drops to deep ``low'' states such as
occasionally shown by BZ~Cam (e.g., Garnavich \& Szkody 1988; Greiner et al.\
2001; Honeycutt et al.\ 2013 and references therein), at least in the ASAS and
ASAS-SN data that we examined.

As typical examples of the star's photometric behavior, Figure~3 shows the
ASAS-SN data for three 100-day intervals between 2017 April and 2018 May. In the
first panel, there are peaks at $V\simeq10.5$ recurring roughly periodically at
intervals of about 14~days. However, in the second panel, the star remained
mostly at its maximum brightness, but with a few drops to $V\simeq11.5$ or
fainter. In 2018 it returned to semi-periodic maxima spaced by about 15.3~days.
There is no evidence for deep eclipses in the data.

It is this quasi-periodic behavior that led to the mis-classification as a
Cepheid in the historical work. The interval between maxima appears to vary, as
indicated by the 11.95-day period reported by Hoffmeister (1956) and the
approximate 10-day timescale found by Kiraga (2012).

\section{Spectroscopic Observations}

In order to study its spectroscopic behavior and determine its orbital period,
we undertook observations of V341~Ara with the SMARTS\footnote{SMARTS is the
Small \& Moderate Aperture Research Telescope System; \url{\tt
http://www.astro.yale.edu/smarts}} 1.5-m telescope at Cerro Tololo Interamerican
Observatory (CTIO), using its RC-focus spectrograph equipped with a CCD camera.
Our observations were made in queue mode by Chilean service personnel, and took
place between 2006 April~7 and July~30. To cover the red region around
H$\alpha$, we used the ``47/Ib'' grating setup, giving a wavelength interval of
5650--6970~\AA\ with a 3-pixel resolution of 3.1~\AA\null. With this setup we
made 17 separate visits to V341~Ara on eight different nights. At each visit we
obtained three individual exposures, usually of 240~s each. 

For the blue region, we used several different setups, depending on the queue
scheduling. These included ``26/I'' covering 3532--5300~\AA\ with a spectral
resolution of 4.3~\AA\ (5 visits), ``56/II'' (4017--4938~\AA, resolution 2.2~\AA;
8 visits), and ``47/IIb'' (4070--4744~\AA, resolution 1.6~\AA; 2 visits). At
each visit we obtained three exposures of 300~s each.

The CCD images were bias-subtracted and flat-fielded, and then the stellar
spectrum was extracted and wavelength-calibrated, all using standard
IRAF\footnote{IRAF is distributed by the National Optical Astronomy Observatory,
which is operated by the Association of Universities for Research in Astronomy
(AURA) under a cooperative agreement with the National Science Foundation.}
routines. Wavelength calibration was accomplished using comparison spectra of
neon (red) or helium-argon (blue) lamps, obtained immediately before and after
each set of three stellar exposures without moving the telescope. 

In Figure~4, we show the ASAS light curve during the interval in 2006 when the
spectroscopic observations were made. At this time the spacing between
photometric maxima was about 16.6~days. Tick marks---red for H$\alpha$, blue for
blue region---indicate the times of our spectroscopic observations, showing that
both the red and blue spectra sampled the high, intermediate, and low states of
V341~Ara. However, a comparison of spectra taken at different brightness levels
showed no clear dependence upon the brightness.

The top panel in Figure~5 shows a blue spectrum created by combining all of the
spectra, and normalizing to a flat continuum. This spectrum is typical of a
nova-like CV, showing broad absorption lines of the Balmer series and
\ion{He}{1}. There are relatively sharp emission lines in the cores of the broad
absorptions. 

The bottom panel of Figure~5 shows the red spectral region, averaged over all of
our spectra. Again we see broad absorption features of hydrogen and \ion{He}{1},
with central emission lines. One curiosity is the apparent presence of
\ion{Si}{2} features in absorption, at 6347 and 6371~\AA, probably with weak
central emission. To our knowledge the \ion{Si}{2} features are not commonly
present in nova-like CVs, although the \ion{Si}{2} doublet near 4130~\AA\ was
seen in absorption in the spectrum of WZ~Sge by Gilliland et al.\
(1986)\footnote{We thank E.~M. Sion for pointing out this paper.} (and may be
present in our spectra as well, although the wavelength is slightly discrepant).
The \ion{Na}{1} doublet also appears to contribute to the broad absorption
feature due to \ion{He}{1} 5875~\AA.

\section{Orbital Period}

The strongest emission line in the spectrum of V341~Ara is H$\alpha$. We
measured the centroid wavelength of the H$\alpha$ emission core in each of our
spectra by fitting a Gaussian profile, using the \textsc{bplot} task in
IRAF\null. The wavelength was converted to radial velocity (RV), and corrected
to the heliocentric frame. Table~1 lists the results.


We used a periodogram analysis to search for a periodic signal in the RVs,
finding a strong peak at about 0.1522~days. This was then refined by computing a
least-squares sinusoidal fit to the RV data, resulting in an ephemeris of ${\rm
HJD} = 2453833.048 \pm0.006 + (0.15216 \pm0.00002)\,E$. The center-of-mass
velocity is $-33.5\,\rm km\,s^{-1}$, and the velocity semi-amplitude is
$32.0\,\rm km\,s^{-1}$.

In Figure~6 we plot the RVs against the orbital phase, along with the sinusoidal
fit. Remarkably, the orbital period of 0.15216~day (3.652~hr) is nearly
identical to the 0.15353 (3.685~hr) period of BZ~Cam (\S1).

\section{Nature of the V341 Ara System}


In this paper we have presented an initial exploration of the nature of the
variable star V341~Ara and its surrounding nebulae. We have confirmed that the
star is a nova-like CV, and we determined its orbital period. 
We consider two possible interpretations of the system:

1. One plausible scenario is that V341~Ara is a CV that is undergoing a chance
high-speed encounter with an interstellar gas cloud, leading to the cloud being
photoionized by ultraviolet radiation from the binary.  In this picture, the
fact that V341 Ara is located well off-center with respect to the nebula Fr~2-11
indicates that the latter was not ejected from the star itself in a previous
nova explosion. The space motion of V341~Ara, of about $80\rm\,km\,s^{-1}$
(including the center-of-mass velocity), is highly supersonic relative to the
gas, presumed to be stationary. It is well established from ultraviolet studies
that the similar system BZ~Cam is ejecting a stellar wind (e.g., Balman et al.\
2014 and references therein). The ``snowplowing'' interaction between a wind
from the accretion disk of V341~Ara and the ISM cloud would produce the
bow-shock region near the star.

2. An alternative picture is that the faint, large nebula Fr~2-11 {\it was\/}
ejected during a nova outburst of V341~Ara in the distant past. Over the past
decade or so, faint nebulae have been discovered around several other known CVs.
These include two members of the Z~Cam subclass (Z~Cam itself, Shara et al.\
2007; AT~Cnc, Shara et al.\ 2012), and a nova-like variable (V1315~Aql, Sahman
et al.\ 2018). In addition, investigations of several faint nebulae have
revealed associations with previously unknown CVs. These include a nova-like
variable in IPHASX J210204.7+471015 (Guerrero et al.\ 2018) and a dwarf nova
(Te~11, Miszalski et al.\ 2016). As we mentioned in \S1, IPHASX J210204.7+471015
is associated with a bow-shock nebula. There is also a weak bow shock on the
southwest side of Te~11. We suspect that the unusually shaped inner [\ion{O}{3}]
nebula of Te~11 may be the product of a similar mass-loss process as experienced
by BZ Cam and V341 Ara, except in this case the lack of a high space motion has
left the nebula relatively less disturbed.

All of these nebulae are approximately centered on the variables and thus almost
certainly are ejecta from unwitnessed classical nova eruptions that occurred
several hundred to a few thousand years ago. However, a particularly instructive
example for our discussion comes from the recent discovery by Shara et al.\
(2017) of a nova shell and a CV associated with the explosion site of the
historical Nova Scorpii AD~1437. The star has been classified as a DQ~Her-type
magnetic CV by Potter \& Buckley (2018). In this case, the star is observed to
lie off-center with respect to the nova shell. Shara et al.\ account for this by
proposing that the ejecta are being decelerated through an interaction with the
ISM, while the remnant star continues its motion. Theoretical investigations of
such scenarios have been made by, e.g., Wareing et al.\ (2007), and indeed can
result in nebulae with an off-center star, and eventually a star completely
outside its ejected nebula.

Consistent with this picture, Shara et al.\ measured the proper motion of the
Nova Sco remnant star, and showed that (assuming the nebula to be stationary)
the star had been located near the center of the nebula in the 15th
Century\footnote{The recent {\it Gaia\/} data release results in a smaller
proper motion for the star than found by Shara et al., but nevertheless in a
direction approximately consistent with their scenario, especially if the nebula
itself has a smaller proper motion in the same direction as the star.}. Using
the proper motion of V341~Ara measured by {\it Gaia\/} (see \S2), and assuming
the H$\alpha$ nebula to be stationary, we find that the star was indeed closest
to the center of the nebula roughly 800~yr ago\footnote{At maximum brightness, a
typical classical nova at a distance of $\sim$150~pc would appear about as
bright as Sirius. Unfortunately, V341~Ara lies so far south that it would not
have been visible to astronomers in Europe, and much of Asia and North
America.}. The appearance of a bow shock in this scenario arises because the
nebula ejected by the nova outburst has been decelerated by interaction with the
ISM, but the star continues its high-velocity motion and ejects a stellar wind
into the surrounding material. Arguing perhaps against this scenario, however,
is the fact that the large nebula shows no pronounced evidence for a
decelerative interaction with the ISM, at least in the material available to us.
Also, if the Fr~2-11 nebula is only 800~yr old, its expansion velocity is about
$220\,\rm km\,s^{-1}$, which should be observationally testable.

%

In conclusion, V341~Ara is one of the brightest nova-like variables, as well as
one of the nearest---and it is a near-twin of the well-studied (but fainter and
more distant) BZ~Cam---yet it has remained largely unknown among the CV
community. We hope that this paper will encourage further studies of this
puzzling system. A useful constraint on its history would come from a
determination of the expansion velocity of the H$\alpha$ nebula Fr~2-11: is the
velocity high, indicating that it was ejected from V341~Ara, or low, indicating
a chance encounter? 

\acknowledgments

We are grateful to David Frew for informing us of his discovery in advance of
publication.

H.E.B. thanks the STScI Director's Discretionary Research Fund for supporting
our participation in the SMARTS consortium, and Fred Walter for scheduling the
1.5-m queue observations. We appreciate the excellent work of the
CTIO/SMARTS service observers, Claudio Aguilera and Alberto Pasten, who obtained
the spectra during many long clear Tololo nights.

B.M. acknowledges support from the National Research Foundation (NRF)
of South Africa.

This research has made use of the services of the ESO Science Archive Facility.


{\it Facilities:} 
\facility{CTIO:1.5m}



\begin{figure}
\begin{center}
\includegraphics[height=4.25in]{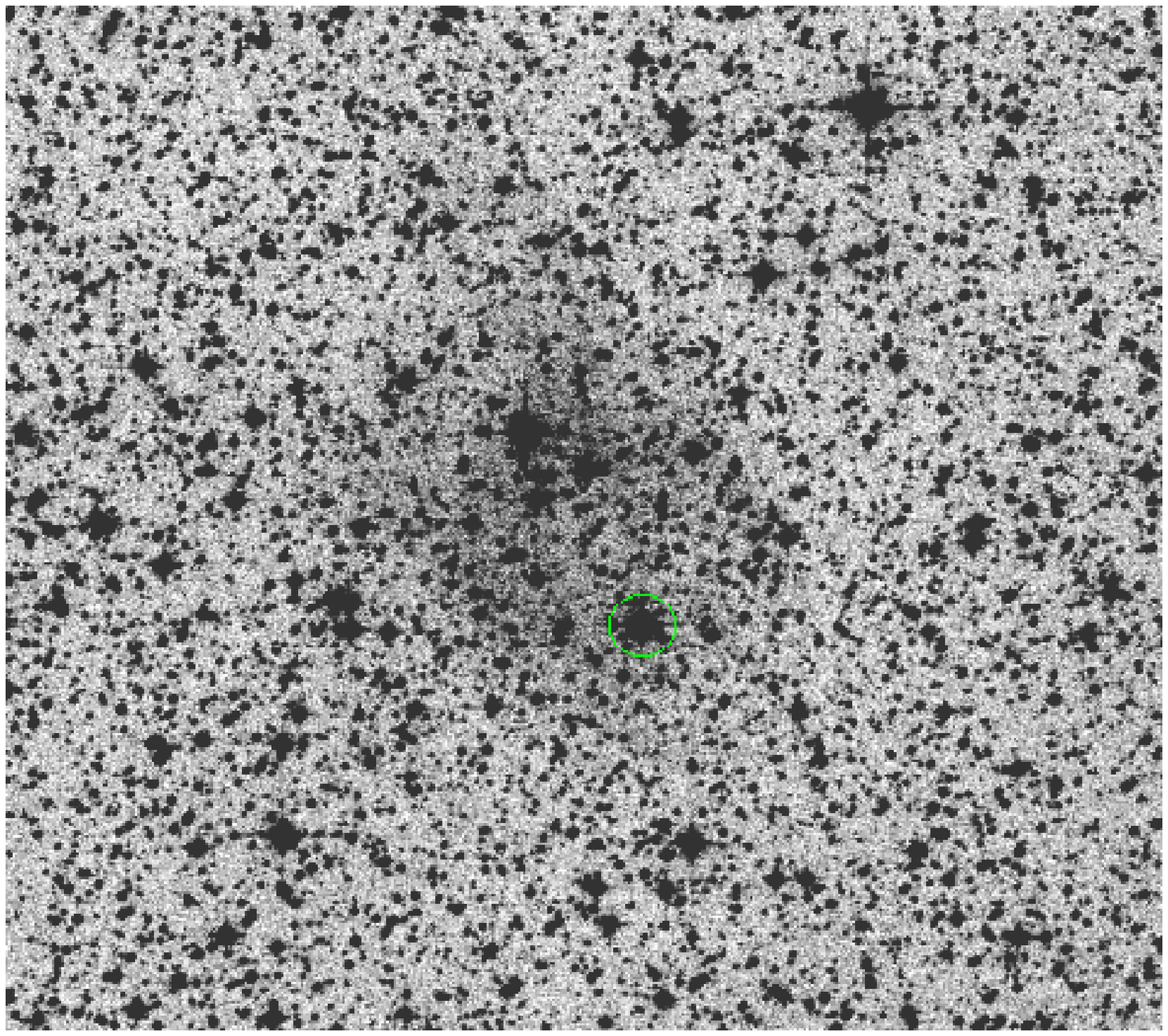}
\figcaption{
Digitized Sky Survey red-sensitive image of the nebula Fr~2-11 associated with
the variable star V341~Ara, marked with a green circle at the south-southwest
edge of the nebula. North is at the top and east on the left, and the height of
the frame is $15'$. The Digitized Sky Surveys were produced at the Space
Telescope Science Institute under U.S. Government grant NAG W-2166.
}
\end{center}
\end{figure}

\begin{figure}
\begin{center}
\includegraphics[height=4.25in]{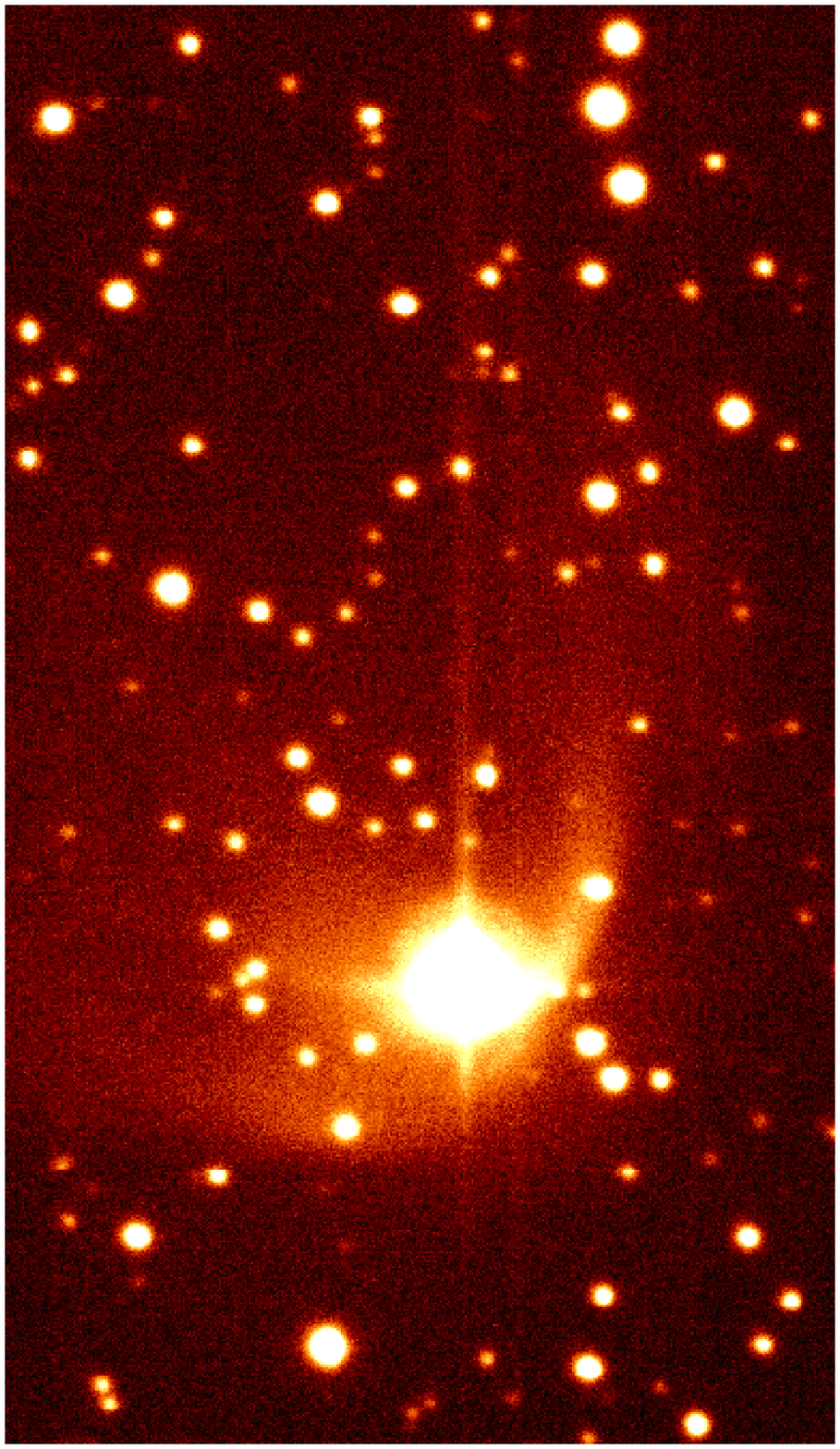}
\hskip0.1in
\includegraphics[height=4.25in]{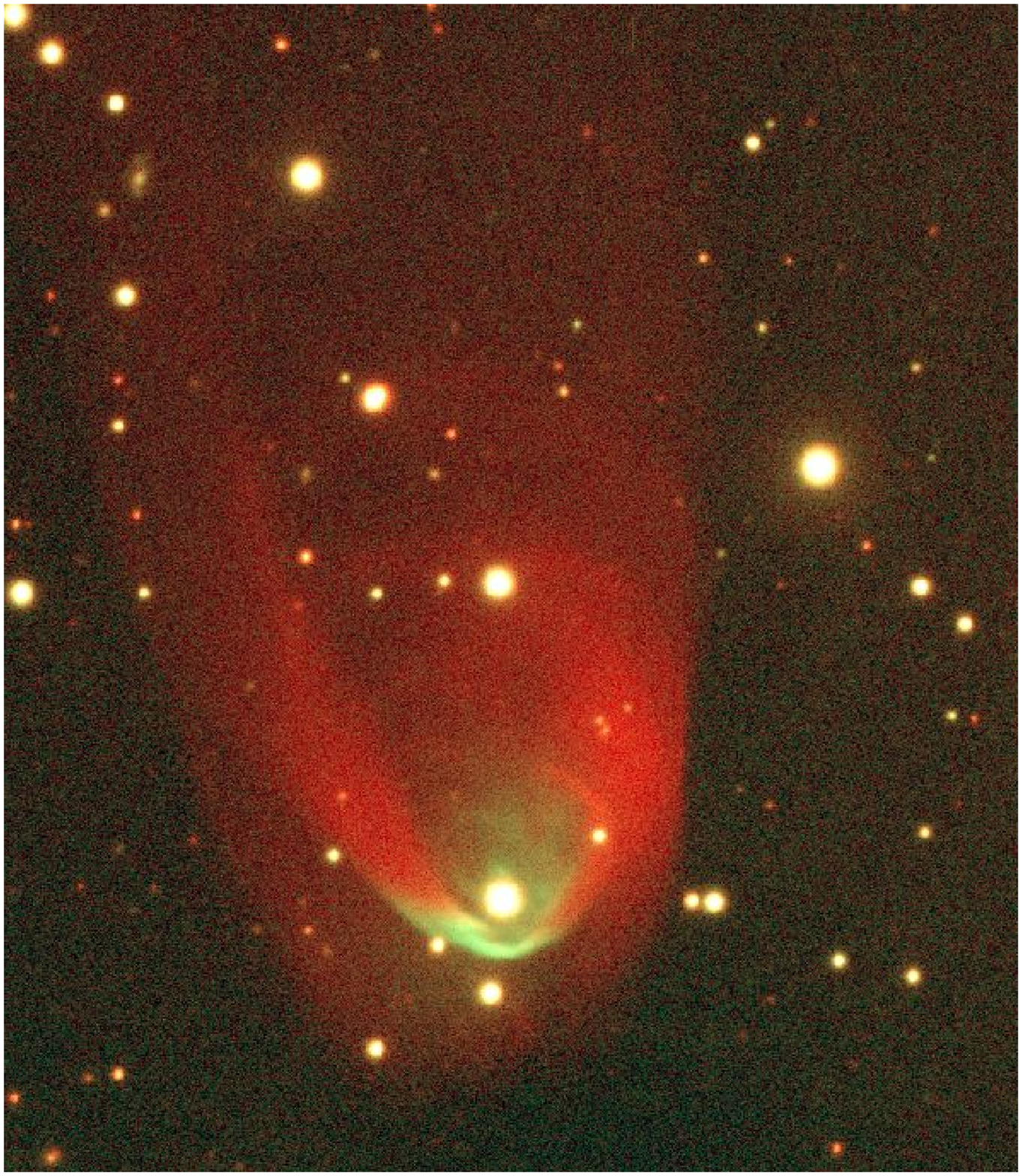}
\figcaption{
False-color renditions of narrow-band emission-line images of two nova-like
variables embedded in bow-shock nebulae. In both frames north is at the top and
east on the left, and logarithmic stretches were used.
{\it Left:} V341~Ara. This is an image in [\ion{O}{3}] 5007~\AA, obtained on
2004 May~27 with the MPG/ESO 2.2-m telescope (see text for details and credits).
Height of frame is $2\farcm9$. The faint vertical features are artefacts.
{\it Right:} The EGB~4 nebula associated with BZ~Cam. This image was obtained
with the Kitt Peak National Observatory Mayall 4-m telescope on 1996 March 13.
Red is H$\alpha$ and green is [\ion{O}{3}]. Height of frame is $6\farcm1$.
}
\end{center}
\end{figure}

\begin{figure}
\begin{center}
\includegraphics[width=4.2in]{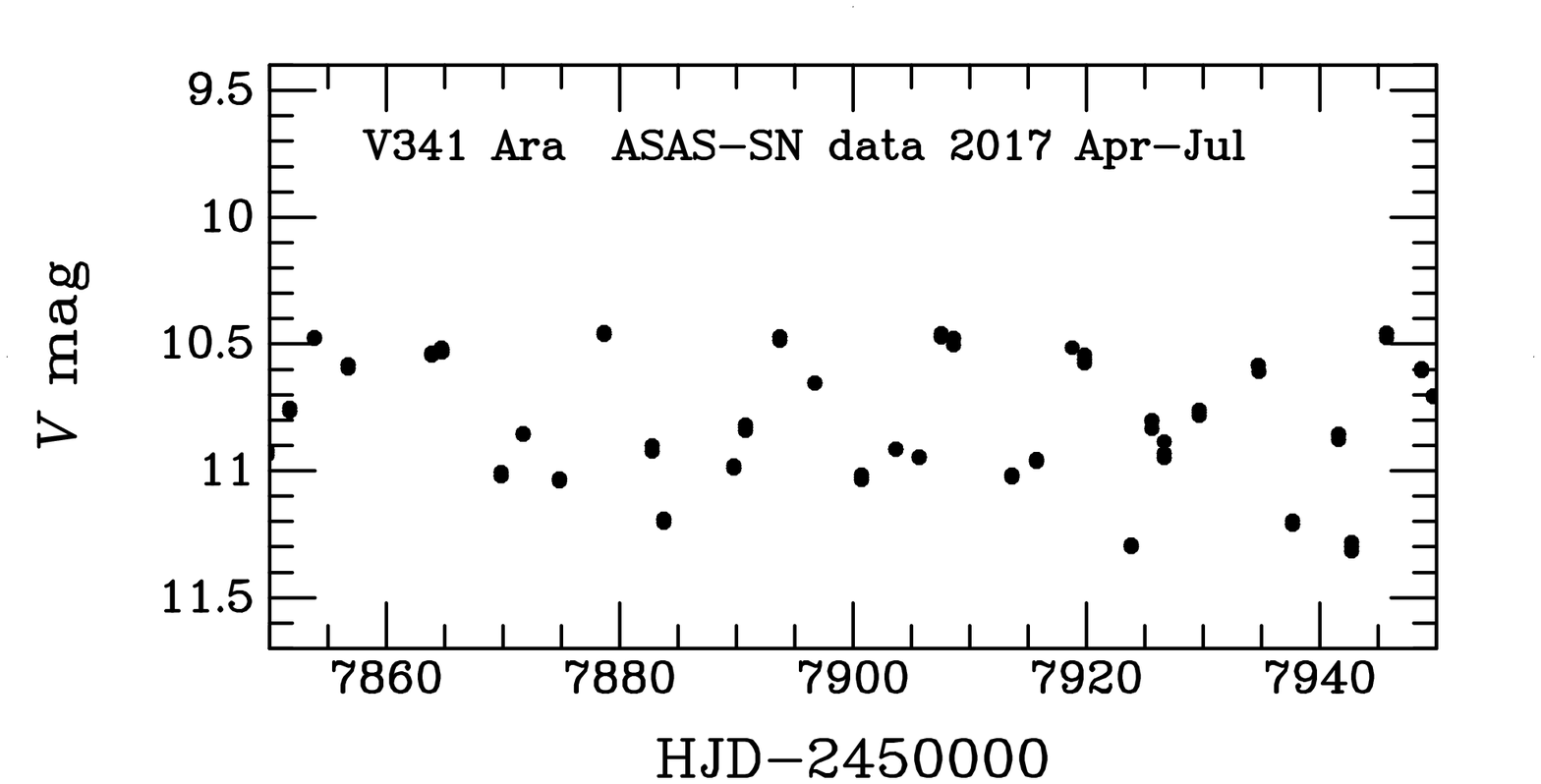}
\vskip0.15in
\includegraphics[width=4.2in]{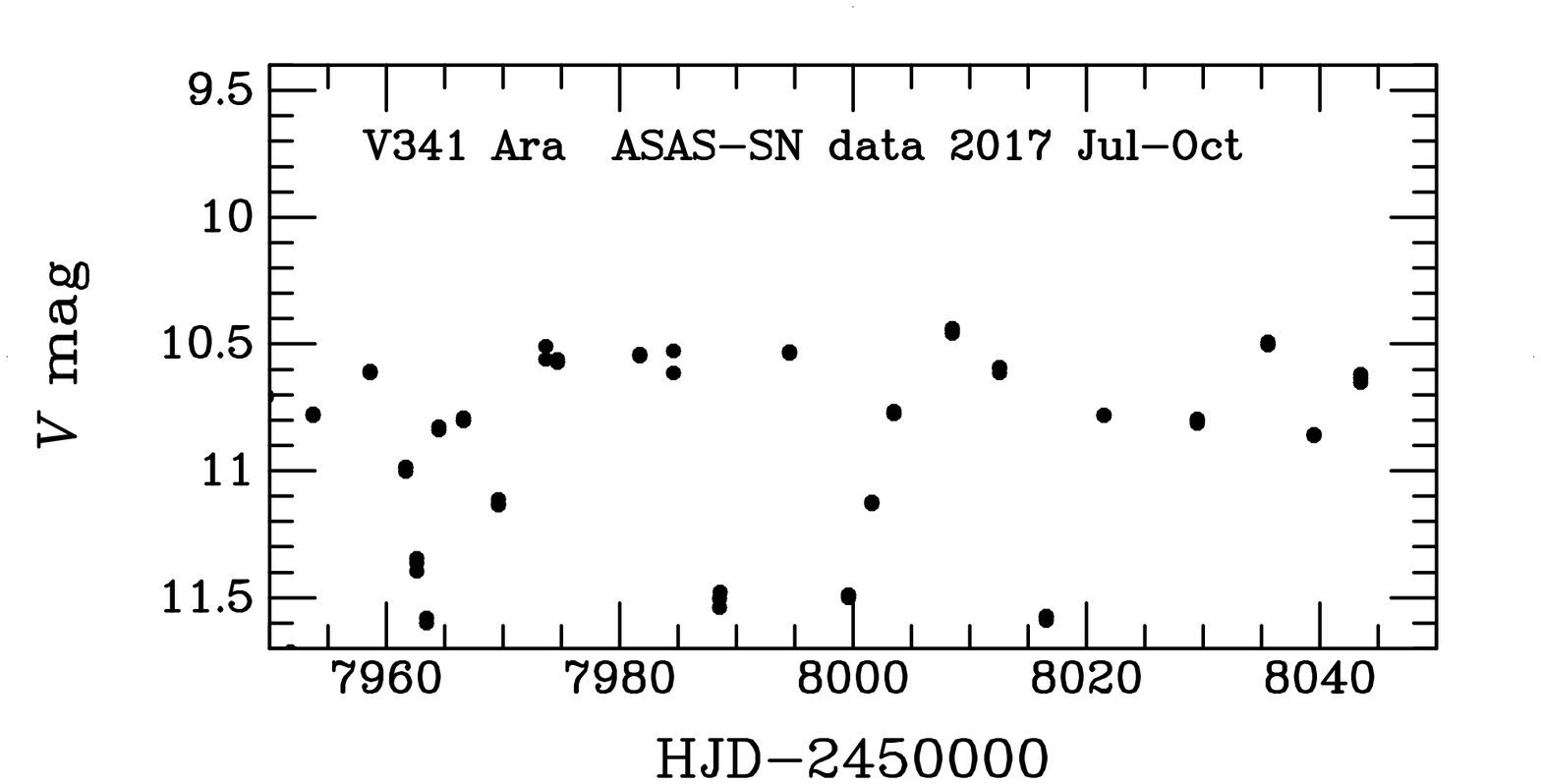}
\vskip0.15in
\includegraphics[width=4.2in]{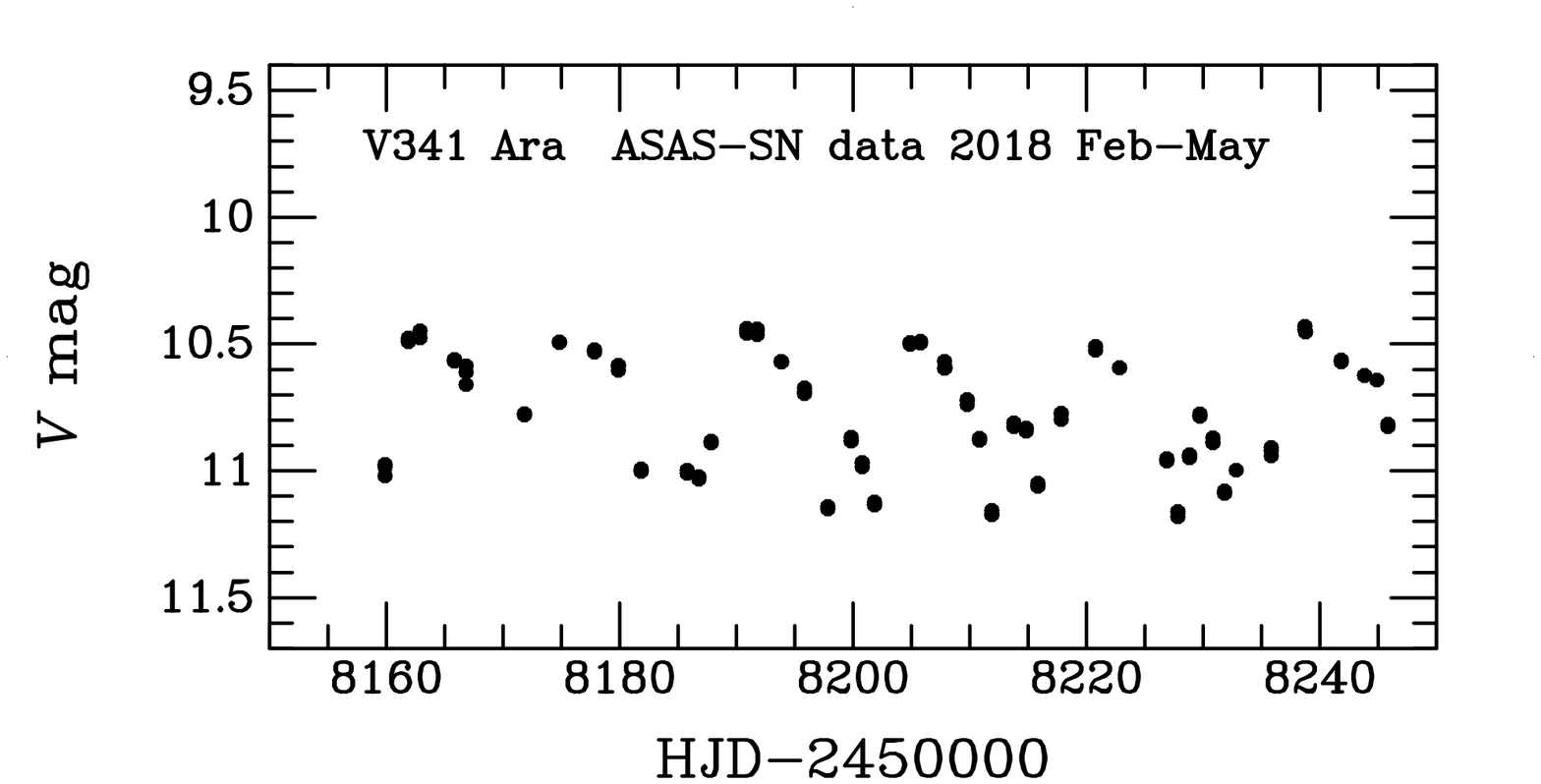}
\figcaption{
ASAS-SN light curves of V341~Ara for three 100-day intervals in 2017--2018.
Magnitude error bars are generally smaller than the plotting symbols. In the
first and third panels, the maxima tended to recur at intervals of about 14--15
days (see text), but in the middle panel panel the star tended to remain bright
with occasional excursions to fainter levels.
}
\end{center}
\end{figure}

\begin{figure}
\begin{center}
\includegraphics[width=5.75in]{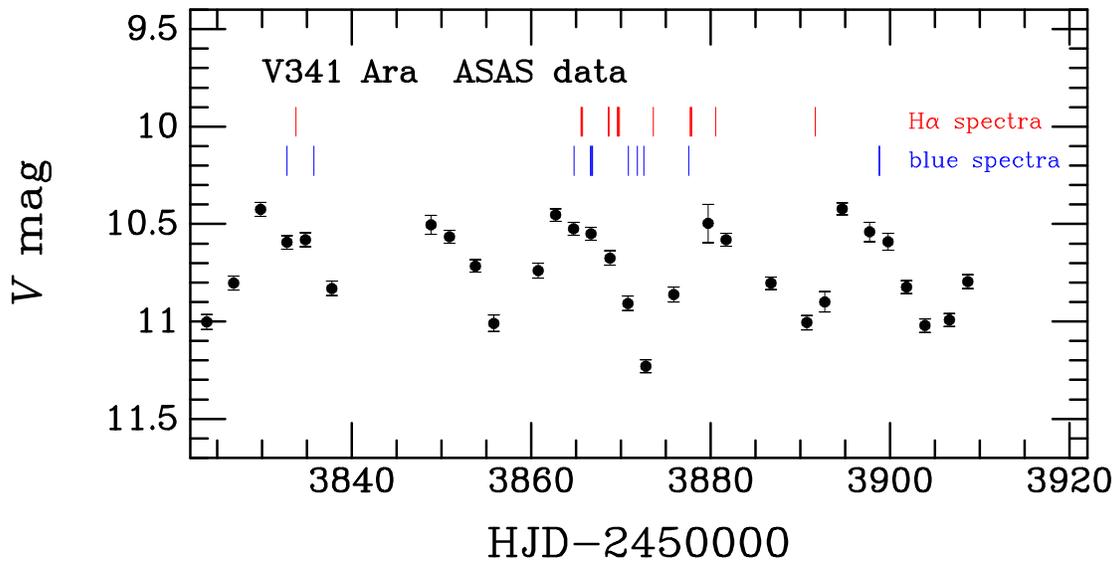}
\figcaption{
ASAS light curve of V341~Ara during the season of our spectroscopic observations
with the SMARTS 1.5-m telescope, 2006 April--July. Red tick marks indicate the
times of H$\alpha$-region spectra, and blue ticks show the times of blue-region
spectra.
}
\end{center}
\end{figure}

\begin{figure}
\begin{center}
\includegraphics[width=5.5in]{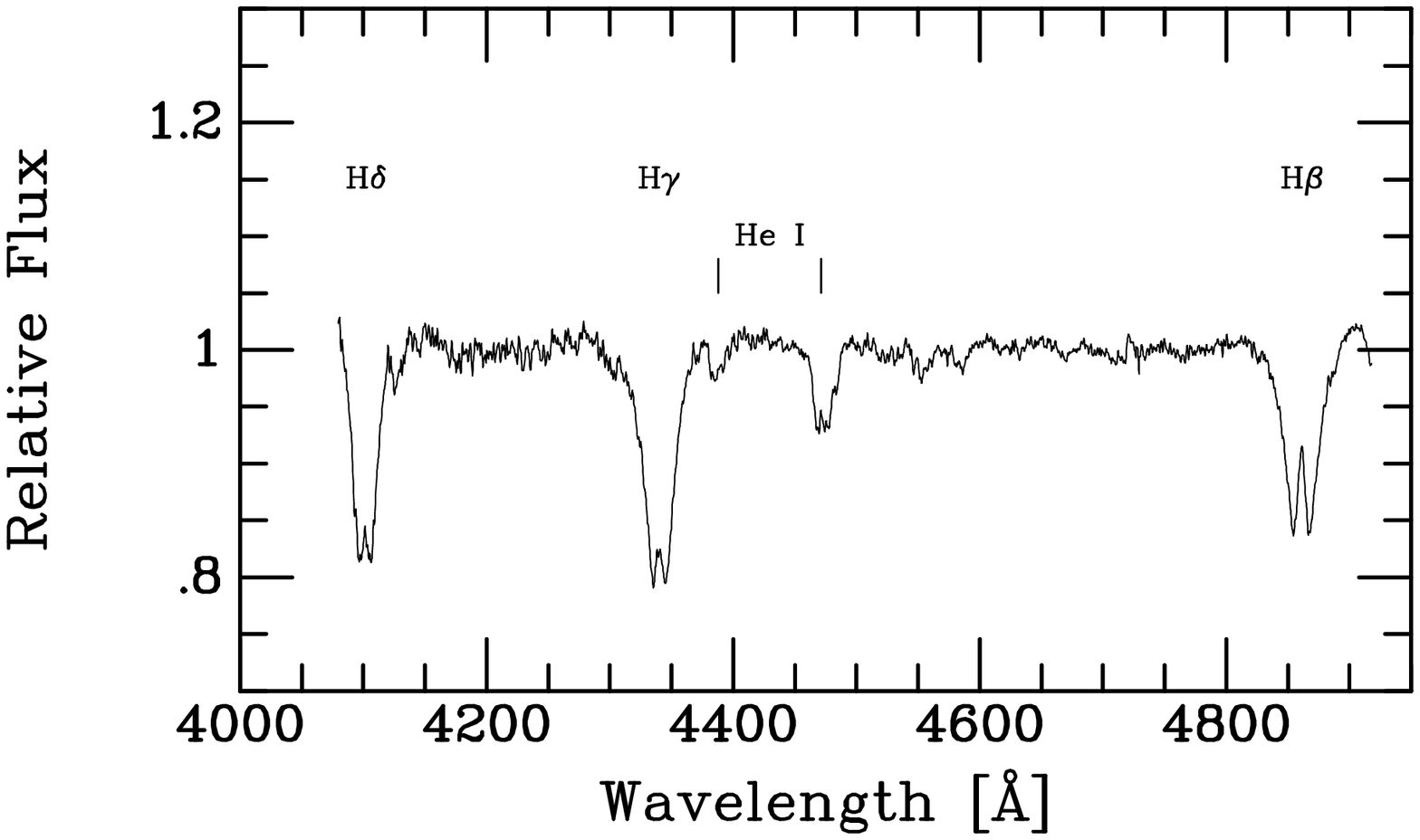}
\vskip0.25in
\includegraphics[width=5.5in]{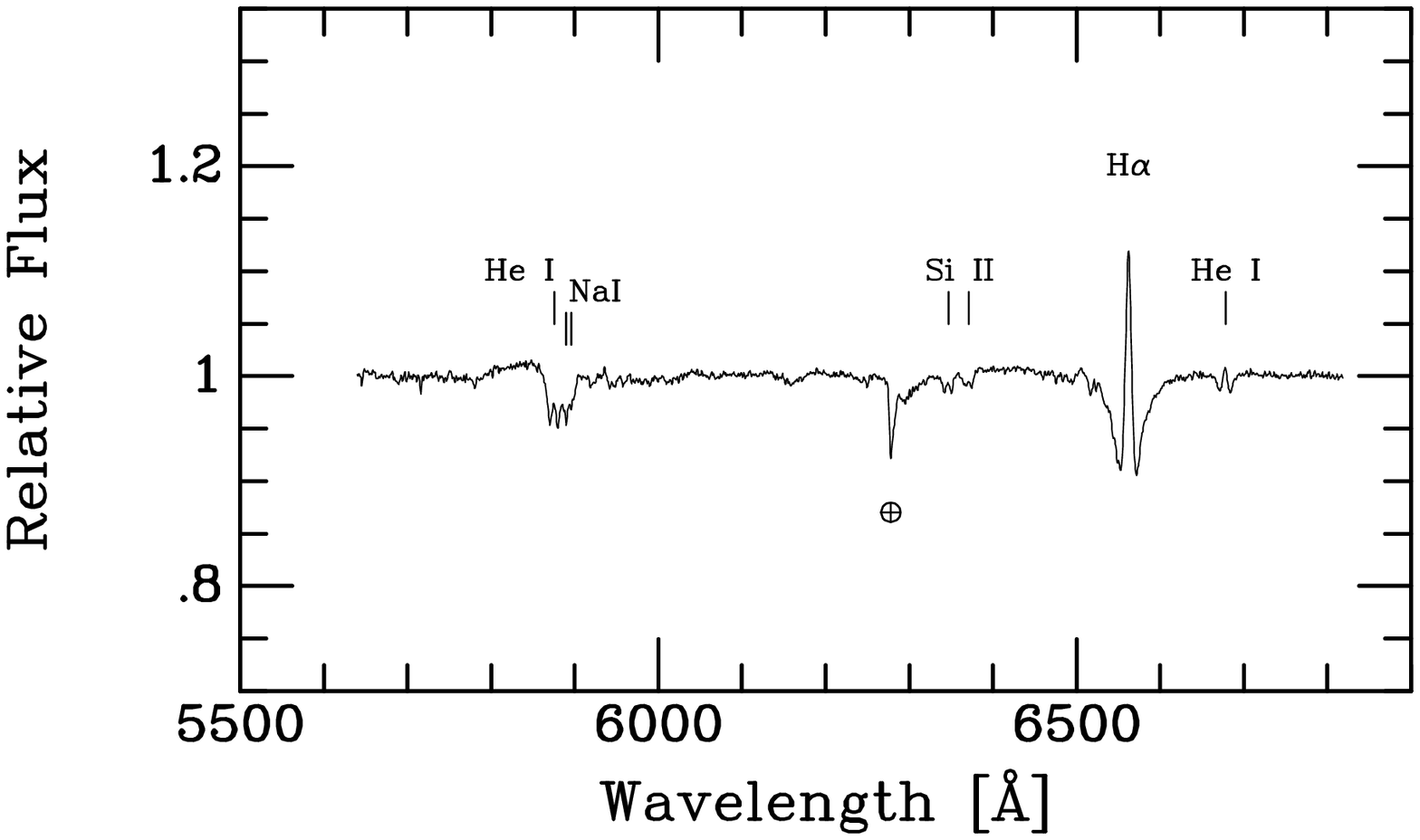}
\figcaption{
Plots of the averages of all of our V341~Ara spectra in the blue (top panel) and
red (bottom panel) wavelength regions, normalized to a flat continuum. The
spectra are typical of nova-like CVs, showing broad absorption features of the
Balmer series and \ion{He}{1}, with central emission peaks. Absorption due to
\ion{Na}{1} and \ion{Si}{2} also appears to be present.
}

\end{center}
\end{figure}

\begin{figure}
\begin{center}
\includegraphics[width=5.75in]{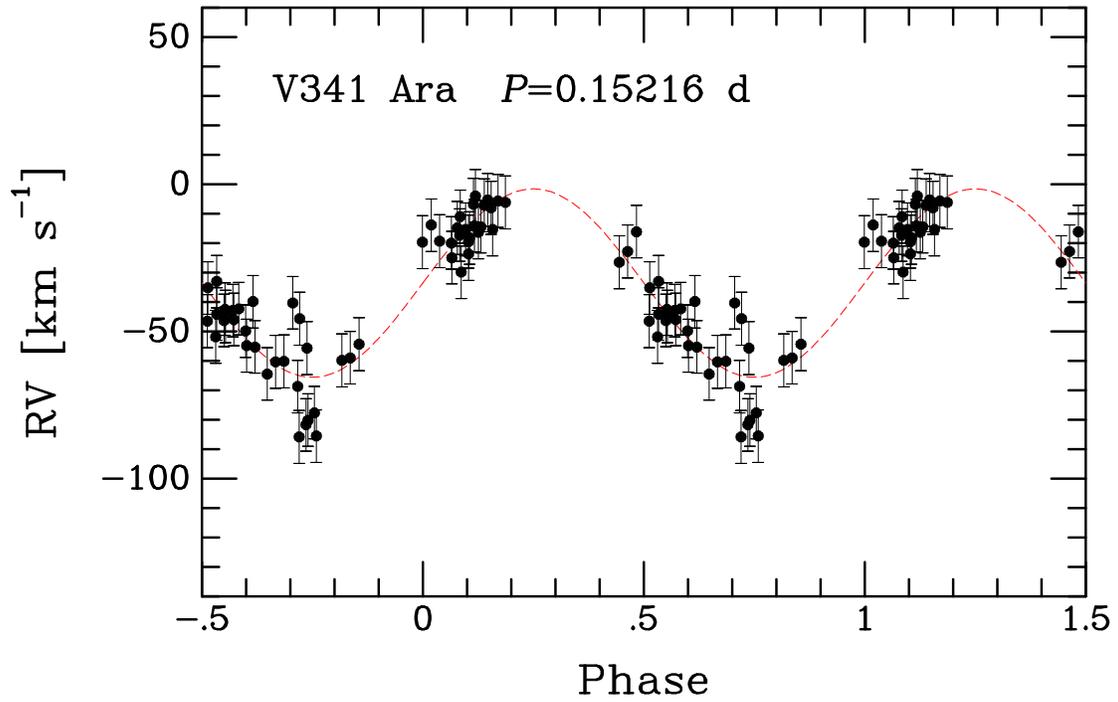}
\figcaption{
H$\alpha$ radial-velocity curve for V341~Ara. The velocities listed in Table~1
are plotted against orbital phase, calculated using the ephemeris ${\rm HJD} =
2453833.048 + 0.15216\,E$\null. The dashed red curve shows a least-squares
sinusoidal fit to the data, with coefficients as listed in the text.
}
\end{center}
\end{figure}

\clearpage

\begin{deluxetable}{lclclc}
\tablewidth{0 pt}
\tablecaption{H$\alpha$ Radial Velocities of V341 Ara}
\tabletypesize{\footnotesize}
\tablehead{
\colhead{HJD$-$2450000} &
\colhead{RV [$\rm km\,s^{-1}$]\tablenotemark{a}} &
\colhead{HJD$-$2450000} &
\colhead{RV [$\rm km\,s^{-1}$]\tablenotemark{a}} &
\colhead{HJD$-$2450000} &
\colhead{RV [$\rm km\,s^{-1}$]\tablenotemark{a}} 
}
\startdata
3833.7553 & $-$64.4 & 3872.7729 & $-$14.8 & 3872.7788 & $-$14.1 \\
3833.7583 & $-$60.4 & 3868.6689 & $-$18.2 & 3873.6027 & $-$51.8 \\
3833.7612 & $-$60.2 & 3868.6719 & $-$16.4 & 3873.6056 & $-$46.3 \\
3865.5671 & $-$85.9 & 3869.5832 & $-$6.9  & 3873.6086 & $-$42.9 \\
3865.5701 & $-$80.1 & 3869.5857 & $-$14.4 & 3877.6978 & $-$26.5 \\
3865.5730 & $-$85.6 & 3869.5882 & $-$5.4  & 3877.7007 & $-$22.7 \\
3865.6279 & $-$3.9  & 3869.6732 & $-$40.3 & 3877.7036 & $-$16.1 \\
3865.6308 & $-$7.2  & 3869.6756 & $-$45.7 & 3877.7821 & $-$19.7 \\
3865.6338 & $-$15.4 & 3869.6781 & $-$55.7 & 3877.7851 & $-$13.9 \\
3865.7187 & $-$68.7 & 3869.7414 & $-$8.0  & 3877.7880 & $-$19.3 \\
3865.7217 & $-$81.7 & 3869.7438 & $-$5.6  & 3877.8635 & $-$44.1 \\
3865.7246 & $-$77.6 & 3869.7463 & $-$6.2  & 3877.8664 & $-$44.9 \\
3868.5787 & $-$46.5 & 3869.8067 & $-$42.3 & 3877.8694 & $-$45.9 \\
3868.5921 & $-$54.8 & 3869.8091 & $-$49.9 & 3880.5310 & $-$25.0 \\
3868.5951 & $-$55.3 & 3869.8116 & $-$39.9 & 3880.5339 & $-$11.0 \\
3868.6250 & $-$59.7 & 3872.6871 & $-$35.2 & 3880.5368 & $-$23.7 \\
3868.6279 & $-$58.9 & 3872.6901 & $-$33.0 & 3891.6384 & $-$19.9 \\
3868.6309 & $-$54.3 & 3872.6930 & $-$42.5 & 3891.6413 & $-$17.3 \\
3868.6660 & $-$29.9 & 3872.7758 & $-$15.3 & 3891.6443 & $-$19.5 \\
\enddata
\tablenotetext{a}{Velocity errors are estimated at about $9.0\,\rm km\,s^{-1}$.}
\end{deluxetable}

\end{document}